\documentclass{amsproc}

\theoremstyle{definition}

\usepackage{graphicx}
\theoremstyle{remark}

\usepackage{subfigure}
\numberwithin{equation}{section}

\newcommand{\bC     }{\mbox{\boldmath$C$}}
\newcommand{\bD     }{\mbox{\boldmath$D$}}
\newcommand{\bA     }{\mbox{\boldmath$A$}}
\newcommand{\bG     }{\mbox{\boldmath$G$}}

\newcommand{\bx     }{\mbox{\boldmath$x$}}
\newcommand{\bzero     }{\mbox{\boldmath $0$}}
\newcommand{\bH     }{\mbox{\boldmath$H$}}
\newcommand{\bPhi     }{\mbox{\boldmath$\Phi$}}
\newcommand{\bS     }{\mbox{\boldmath$S$}}
\newcommand{\bL     }{\mbox{\boldmath$L$}}
\newcommand{\bI     }{\mbox{\boldmath$I$}}

\begin{document}

\title{On the spectra of large sparse graphs with cycles}

\author{D. Boll\'e}
\address{Instituut voor Theoretische Fysica, KULeuven, Celestijnenlaan 200D,
B-3001 Leuven, Belgium}
\email{desire.bolle@fys.kuleuven.be}

\author{F. L. Metz}
\address{Dipartimento di Fisica, Sapienza Universit\`a di Roma, Piazzale A. Moro
2, 00185 Roma, Italy
}
\email{fmetzfmetz@gmail.com}

\author{I. Neri}
\address{Universit\'e Montpellier 2, Laboratoire Charles Coulomb UMR
5221, F-34095, Montpellier, France }
\email{izaakneri@gmail.com}


\date{MSC2010}

\dedicatory{This paper is dedicated to Fritz Gesztesy.}


\begin{abstract}
We present a general method for obtaining the spectra of  large graphs with
short cycles using ideas from statistical mechanics of disordered systems.  This
approach leads to an algorithm that 
determines the spectra of graphs up to a high accuracy.  In particular, for
(un)directed regular graphs with cycles of arbitrary length we derive 
exact and simple equations for the resolvent of the associated adjacency
matrix.  Solving these equations we obtain analytical formulas  for the spectra
and the boundaries of their support. 
\end{abstract}

\maketitle

\section{Introduction}

In spectral graph theory one  uses spectral analysis to study the
interplay between the  topology of a graph and  the dynamical processes modelled
through the graph \cite{Mohar2, Chung, Brouwer, Mohar1}.  In fact, various
 dynamical processes  in disciplines ranging from physics, biology, information
theory, and chemistry to technological and social sciences are modelled with
graph theory. Therefore, graph theory forms
a unified framework for their study \cite{Bar, New}.  In
particular, one associates a certain matrix to a graph (e.g.~the adjacency matrix, the
Laplacian, the google matrix) and studies the connection between  the spectral properties
of this matrix and  the properties of the dynamical processes governed through
them. We mention some  studies in this context: the stability of
synchronization processes \cite{Pec, Wilk}, the robustness and the
effective resistance of networks \cite{Klein}, error-correcting codes
\cite{Hoory}, etc.

These examples illustrate how  spectral graph theory relies  to a large extent
on the capability of determining spectra of sparse graphs.
 It is thus important to develop mathematical methods which allow to derive
in a systematic way  exact analytical as well as numerical
results on spectra of large  graphs.  For an overview of analytical
results  on the spectra of infinite graphs we refer to the paper of Mohar and Woess
\cite{Mohar1}.  

Recently, the development of exact results for large sparse graphs
has been reconsidered using ideas from statistical physics of disordered
systems.  In this approach one formulates the spectral analysis of graphs in a
statistical-mechanics language using a
complex valued measure \cite{EA}. The spectrum
is given as the free energy density of this measure, which can be calculated
using methods from disordered systems such as the
replica method \cite{Kuhn}, the cavity method
\cite{Rogers1, Rogers2} or the super-symmetric method \cite{Fyod}.   This
approach is exact for infinitely large graphs that do not contain cycles of
finite length.  In recent works the cavity
method has been generalized to the study of spectra of graphs with a
community structure
\cite{Rogers3} and  spectra of small-world graphs \cite{Mour}. The replica
and cavity methods 
have also been used to derive the largest eigenvalue of sparse random
matrix ensembles \cite{Kab}.
Although the cavity method is
heuristic, it has been considered in a rigorous setting for undirected sparse
graphs using the theory of local weak convergence ~\cite{Bordenave}.
 However, for directed sparse graphs
the asymptotic convergence of the spectrum to the resultant cavity expressions
has not been shown \cite{BordChaf}.   Nevertheless, recent studies have
shown the asymptotic convergence of the spectrum  of highly
connected sparse matrices to the circular law \cite{Tao1, Tao2, Go, Wood} and
have proven the asymptotic convergence of the spectrum of  Markov generators
on higly connected random graphs \cite{BordenaveTT}. 
These studies, however, do not concern finitely connected graphs. 

In this work we extend the statistical-mechanics formulation, in particular the
cavity method for the spectra of large sparse graphs which are locally 
tree-like \cite{Rogers1, Rogers2, Bordenave}, to  large sparse graphs
with many short cycles. Such an extension is 
relevant because cycles do appear in many real-world systems such as the internet
\cite{Gleiss, Bianc1, Bianc2}.  We derive a set of resolvent equations for
graph ensembles that contain many cycles of finite length. These equations are exact
for infinitely large (un)directed Husimi graphs \cite{Husimi, Har} and solving them
constitutes an algorithm for determining the spectral density of these graphs.  

First, we show how this algorithm determines the
spectra of irregular Husimi graphs up to a high accuracy, well corroborated by
numerical simulations.  Then we derive novel analytical results not only for the
spectra of undirected regular Husimi graphs, but also for the spectra of
directed regular Husimi graphs.  In particular, we show that the boundary of the
spectrum of directed Husimi graphs composed of cycles of length $\ell$, is
determined by a hypotrochoid with radii $R/r = \ell$ ($R$ being the radius of
the fixed circle and $r$ of the moving circle).  Not many analytical
expressions for the spectra of directed sparse
random graphs and non-Hermitian random matrices are known (besides some
exceptions, see \cite{Neri2012}).   A short
account of some of our results on regular graphs has appeared in
\cite{Metz2011}.

\section{Ensembles of graphs}
Networks appearing in nature are usually modelled with theoretical ensembles of
graphs. These  ensembles  consist of randomly constructed graphs  with  certain
topological constraints on their connectivity.  Model systems allow for a better 
understanding of the properties of the more complex
real-world systems \cite{Bar, New}.

In this work we consider simple graphs
$\mathcal{G} = (V,E)$ of size $N = |V|$ consisting of a discrete set of vertices $V$ and  a set of
edges $E\subset V\times V$.  
Simple graphs are uniquely defined by their adjacency
matrix $\bA$, with elements $\left[\bA\right]_{ij} = A_{ij}\in
\left\{0,1\right\}$ for  $i,j\in V$ and $\left[\bA\right]_{ii} = 0$ 
for $\forall \,\, i$.  We have
$A_{ij}=1$ when $(i,j)\in E$ and
zero otherwise.  An  undirected
edge is present between $i$ and $j$ when 
$A_{ij}=A_{ji}=1$, a directed edge is present from $i$ to $j$ when
$A_{ij}=1$ and $A_{ji}=0$, while
$A_{ij}=A_{ji}=0$ indicates that there is no edge present between $i$ and $j$.

We define graph ensembles through a
normalized distribution $P(\bA)$ for the adjacency matrices $\bA$.  
Selecting a graph from the ensemble corresponds with drawing randomly 
an adjacency matrix from the distribution $P(\bA)$. We 
always consider ensembles of infinite graphs, for which $N\rightarrow \infty$. 
This is implicitly assumed throughout the whole paper.
Below, we first define random graphs which are locally tree-like.  Their
spectral properties have been considered in several studies
\cite{Mohar1, Rogers1, Rogers2, Bordenave, Mckay, Khor}. Second, we define
ensembles of graphs with many cycles: the cacti or Husimi graphs \cite{Husimi,
Har}.    
\begin{figure}[h!]
\begin{center}
\includegraphics[angle = 0, scale=0.2]{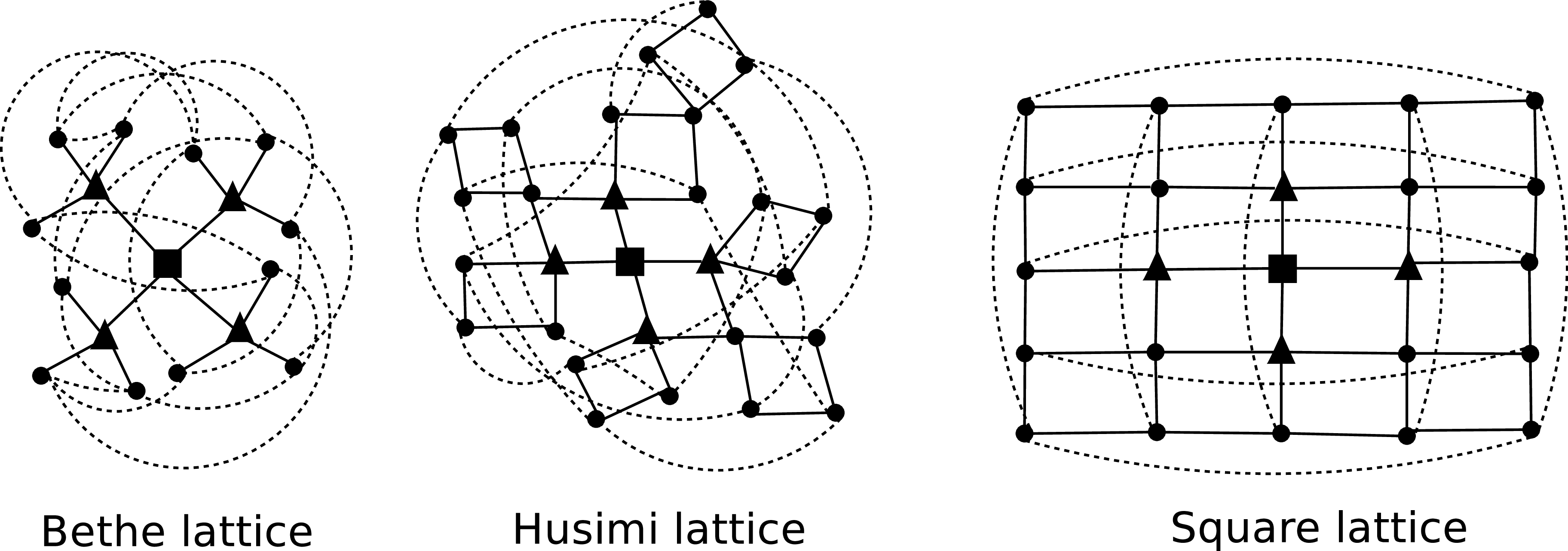}
\caption{Three different regular graphs rooted around the
central vertex.  The central vertex is denoted by a square while its neighbours
are denoted by triangles.  The dashed lines indicate the edges between
the leaves of the rooted graph.  Left: a $4$-regular graph. The rooted graph is
locally tree-like.  Center: a $(4,2)$-regular Husimi graph. The rooted graph is not
tree-like but does maintain an infinite-dimensional nature.  
Right: a non-random graph (the square lattice).}
\label{cavities1A}
\end{center}
\end{figure}  
\subsection{Graphs with a local tree structure}
Ensembles of random graphs with only certain constraints on the vertex degrees 
are locally tree-like.  Some well-studied examples with these characteristics
are the ensemble of $c$-regular  graphs (also called Bethe lattices) and the
ensemble of irregular Erd\"os-R\'enyi  graphs
\cite{Bol}. For undirected graphs, these ensembles can be formally defined as
follows:
\begin{itemize}
\item  $c$-regular graphs with fixed connectivity $c$:
\begin{equation}
 P_{\rm reg}(\bA; c) \sim \prod^N_{i=1}\delta(c_i; c) \prod_{i<j}\delta(A_{ij};
A_{ji}),\label{eq:ensemble1}
\end{equation}
with $c_i = \sum_j A_{ji}$ the degree of the $i$-th vertex. Figure
\ref{cavities1A} presents a sketch of a Bethe lattice.
\item Erd\"os-R\'enyi graphs with mean connectivity $c$:
\begin{equation}
 P_{\rm irreg}(\bA; c) \sim
\prod_{i<j}\left(\frac{c}{N}\delta\left(A_{ij};1\right) +
\left(1-\frac{c}{N}\right)\delta\left(A_{ij};0\right)\right) 
\prod_{i<j}\delta\left(A_{ij};A_{ji}\right). \label{eq:ensemble2}
\end{equation}
In this ensemble the  degrees fluctuate from vertex to vertex within the graph. 
The distribution of vertex degrees converges to a
Poissonian distribution with mean $c$ in its asymptotic limit $N\rightarrow
\infty$. 
\end{itemize}

Ensembles of directed graphs can be defined similarly by leaving out the 
symmetry constraints in the definitions
eqs~(\ref{eq:ensemble1}) and (\ref{eq:ensemble2}) and by taking into
consideration the difference between $A_{ij}$ and $A_{ji}$.  
For instance, we define a $c$-regular ensemble of directed graphs by the
constraint $c^{\rm out}_i = c^{\rm in}_i = c$ for any $i$, with the indegree
defined as $c^{\rm in}_i \equiv \sum_j A_{ji}$
and the outdegree as $c^{\rm out}_i \equiv \sum_j A_{ij}$.  

The Erd\"os-R\'enyi and the $c$-regular ensemble have the local tree property,
i.e.~the probability to encounter a cycle of finite
length in the local neighbourhood of a vertex becomes arbitrary small for
$N\rightarrow\infty$ \cite{Bol}. The
local tree property of random graphs is an  important characteristic  which
has allowed to determine the spectra of
sparse graphs in previous works \cite{Rogers1, Rogers2, Bordenave}.    This
property is illustrated in figure \ref{cavities1A} for the Bethe lattice: cycles
only appear at the leaves of the tree, where the vertex constraints have to be 
satisfied such that their typical length is of the order $\log(N)$. 
\subsection{Ensembles with cycles}
\begin{figure}[h!]
\begin{center}
\includegraphics[angle = 0, scale=0.3]{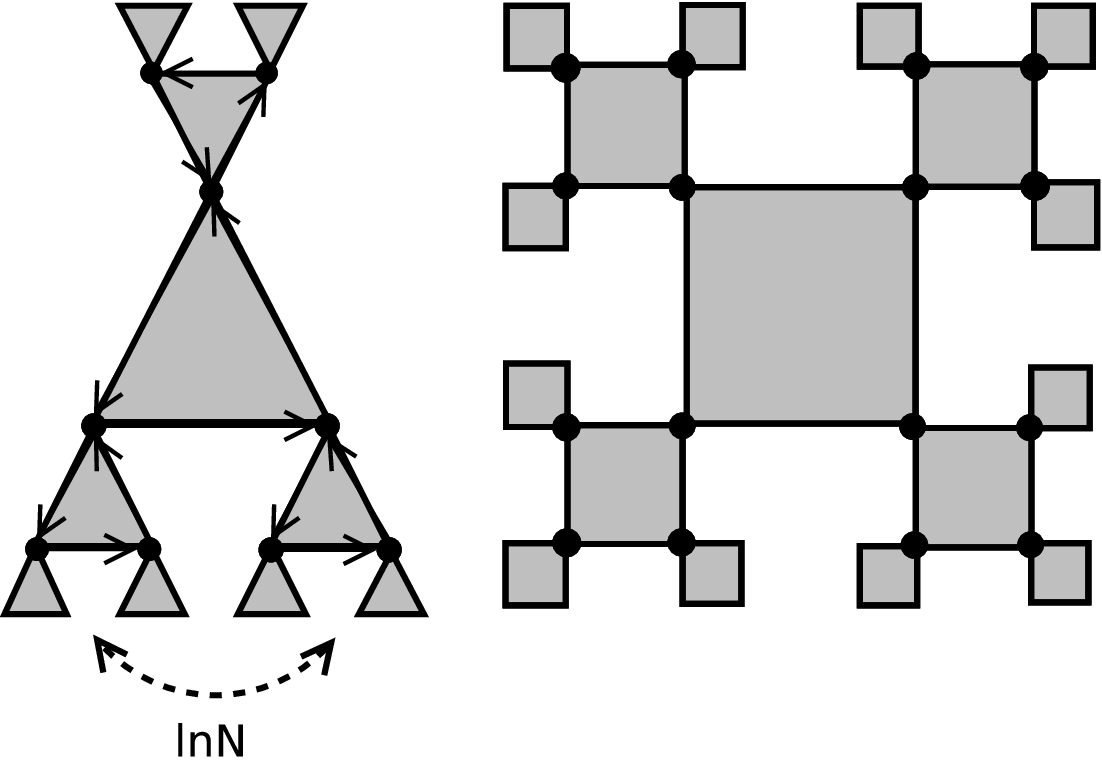}
\caption{\cite{Metz2011} Local structure of some $(\ell,c)$-regular
Husimi graphs, with $\ell$ the cycle length and $c$ the cycle connectivity.  
Left: directed $(3,2)$-regular Husimi graph. Right:
undirected $(4,2)$-regular Husimi graph. The Husimi graphs are infinite
dimensional as different branches of a central node are separated by a
distance of the order $\mathcal{O}(\ln N)$.}
\label{cavities1}
\end{center}
\end{figure}

Next, we define graph ensembles which  contain many short cycles and 
are, therefore, not locally tree-like.  A cycle of length $\ell$ is defined by
 a set of nodes $(i_1, i_2, \cdots
i_\ell)$ with an edge connecting each pair $(i_{n},
i_{(n+1)({\rm mod} \:\ell)})$, for
$n=1,\ldots,\ell$. 
The set of all $\ell$-tuples is denoted
by $V^\ell$.  We define  the following
ensembles which are very similar to
the Husimi graphs in  \cite{Husimi, Har}: 
\begin{itemize}
\item $(\ell, c)$-regular Husimi graphs with fixed connectivity $c$ and
fixed cycle length~$\ell$: 
\begin{eqnarray}
  P_{\rm reg}(\bA; c, \ell) & \sim & \prod^N_{i=1}\delta(c_i; c)
\prod_{i<j}\delta(A_{ij};
A_{ji}),
\end{eqnarray}
where $c_i$ denotes now the number of $\ell$-cycles of $\bA$ 
incident to the $i$-th vertex.   In figure \ref{cavities1} we show the 
typical local neighbourhood of an undirected (4,2)-regular Husimi graph.
\item irregular Husimi graphs with fixed cycle length $\ell$ and mean
loop connectivity $c$:
\begin{eqnarray}
 \lefteqn{ P_{\rm irreg}(\bA; c, \ell) \sim
\prod_{(i_1,i_2,\ldots,i_{\ell})\in V^\ell}\left[\frac{c}{N^{\ell-1}}
\, \delta\left(A_{i_1 i_\ell}\prod^{\ell-1}_{k=1}A_{i_{k}i_{k+1}}
;1\right)
\right.}&& 
\\
&&\left.
+\left(1-\frac{c}{N^{\ell-1}} \right)\delta\left(A_{i_1 i_\ell}
\prod^{
\ell-1 } _ { k=1}A_{i_{k}i_{k+1}}
;0\right)\right] \times \prod_{i<j}\delta(A_{ij};
A_{ji}) 
\nonumber \\
&& \times \prod_{(i,j)}\delta\left(A_{ij}\sum_{(i, k_1,k_2,\cdots,
k_{\ell-2}, j)\in V^{\ell}_{ij}}A_{ik_1}\left(\prod^{\ell-3}_{n=1} A_{k_n
k_{n+1}}\right) A_{k_{\ell-2}j}; A_{ij} \right).\nonumber \label{eq:Poiss3}
\end{eqnarray}
The set $V^\ell_{ij}\subset V^\ell$ consists of all $\ell$-tuples $(i, k_1,
\cdots, k_{\ell-2}, j)\in V^{\ell}$.
The first factor gives a weight
$c/N^{\ell-1}$ to each randomly drawn cycle of length $\ell$, the 
second factor is the constraint which makes the graph undirected and the last
factor is a constraint which takes into consideration that each edge must
belong to exactly one cycle of length $\ell$.
\end{itemize}
We remark that in our notation of Husimi graphs the mean cycle connectivity $c$
is the average number of cycles of length $\ell$ connected to a certain 
vertex, while $2c$ denotes the mean degree of each vertex. Note that 
for $\ell=2$ the ensembles of Husimi graphs introduced here reduce to
the corresponding ensembles of random graphs which are locally tree-like
(discussed in the previous subsection).

As before, we can define as well
directed ensembles of Husimi graphs by removing the symmetry
constraint $\prod_{i<j}\delta(A_{ij}; A_{ji})$ and taking into consideration
the possible difference beween $A_{ij}$ and $A_{ji}$.  For instance, for a
regular Husimi graph we set the mean vertex indegree and the mean vertex
outdegree equal to $2c$. In figure \ref{cavities1} we present the 
neighbourhood of a vertex in a directed (3,2)-regular Husimi graph. 

Husimi graphs are not locally tree-like since they are composed of cycles,
but they do have an infinite-dimensional character.  Indeed, the local
neighbourhood of a vertex contains different branches which are only
connected by cycles of order $\log(N)$, see figures \ref{cavities1A} and \ref{cavities1}. 
This property allows us to present an exact spectral analysis.

\section{Spectral analysis of undirected graphs with cycles}
In this section we show how to determine the spectral properties of undirected
graphs with cycles. For regular graphs,  our approach, briefly presented in
\cite{Metz2011}, consists of four steps.  
First, we  write the spectrum in statistical mechanics terms using a
complex measure.   Then we use the infinite-dimensional
nature of the graphs to derive a closed expression in the submatrices 
of the resolvent. Next, we simplify the resolvent
equations to the specific case of regular undirected Husimi graphs. 
Finally, we solve the resultant algebraic expressions to find analytical
expressions for the spectra of regular Husimi graphs. 
\subsection{Statistical mechanics formulation of the spectrum}
We define the resolvent of a real symmetric matrix  $\bA$ as 
\begin{eqnarray}
 \bG(z) \equiv \left(z \bI_{N}-\bA \right)^{-1},
\end{eqnarray}
where $\bI_{N}$ is the $N \times N$ identity matrix and 
$z\in \mathbb{C}$, with $\rm{Re}(z) = \lambda$ and $\rm{Im}(z)< 0$.  From the
resolvent we can determine
the asymptotic density $\rho(\lambda)$ of the real eigenvalues $\lambda$ of an
ensemble of symmetric matrices
$\bA$ \cite{Mohar1}: 
\begin{eqnarray}
 \rho(\lambda) = \lim_{\rm{Im}(z)\rightarrow 0^-}\lim_{N\rightarrow \infty}\frac{1}{\pi N}
{\rm Im}\left[{\rm Tr}\bG\right]. 
\label{eq:specRes}
\end{eqnarray}
We have left out ensemble averages since we assume that the spectrum self-averages
in the limit $N\rightarrow\infty$  to a deterministic value.

A common method in statistical physics is to write  the
resolvent elements and the spectrum  as averages over a certain complex-valued
measure $P(\bx;z)d{\bx}$ \cite{EA}. Indeed, when defining the normalized Gaussian function
\begin{eqnarray}
 P(\bx;z) \sim  \exp\left(-\frac{i}{2}
\bx^T\bG^{-1}(z)\bx\right),
\end{eqnarray}
the diagonal elements $\left[\bG\right]_{ii} = G_{i}$ are given by
\cite{Metz2010}
\begin{eqnarray}
 G_{i}  = i\langle x^2_i\rangle, 
\end{eqnarray}
with $\langle \cdot \rangle$ the average with respect to $P(\bx;z)$.
In this formalism the spectrum (\ref{eq:specRes}) is written as \cite{Kuhn}
\begin{eqnarray}
 \rho(\lambda) =  -\lim_{\rm{Im}(z)\rightarrow 0^-}\lim_{N\rightarrow \infty} 
\frac{2}{\pi N} {\rm Im} \left[ \frac{\partial}{\partial \lambda} \ln
\mathcal{Z} \right]  \, , \label{eq:rhoenergy}
\end{eqnarray}
with 
\begin{eqnarray}
 \mathcal{Z} = \int d\bx  \exp\left(-\frac{i}{2}
\bx^T\bG^{-1}(z)\bx\right).
\end{eqnarray}
such that
\begin{eqnarray}
\rho(\lambda) &=& \lim_{\rm{Im}[z]\rightarrow 0^-} 
\frac{1}{\pi N} \sum^N_{i=1}{\rm{Im}}G_i\left(z\right), \quad {\rm where} \quad \ \lambda =
\rm{Re}[z]. \label{eq:spectrumRec}
\end{eqnarray}

Hence, in the language of statistical mechanics the spectral density $\rho(\lambda)$ of eigenvalues directly related to the
imaginary part of the derivative of the free energy density with respect 
to $\lambda$, as is seen explicitly in eq.(\ref{eq:rhoenergy}).
\subsection{The resolvent equations} 
We employ the cavity method to determine the spectral properties
of locally tree-like graphs \cite{Rogers1, Bordenave}. 
We construct a matrix $\bA^{(i)}$ (and its corresponding resolvent
$\bG^{(i)}$) through deletion of the $i$-th column and the $i$-th
row in $\bA$.  The graph $\mathcal{G}^{(i)}$ induced by
the matrix $\bA^{(i)}$ is usually referred to as the
cavity graph \cite{Mezard}. On this graph we use the
Bethe-Peierls tree approximation \cite{Mezard, Bethe} to arrive at 
the following set of resolvent equations:
\begin{equation}
G^{(j)}_i = \frac{1}{z - \sum_{k\in \partial_i\setminus
j}A^2_{ik}G^{(i)}_k}, \,\,\,\,\,
 G_i = \frac{1}{z - \sum_{k\in \partial_i
}A^2_{ik}G^{(i)}_k},  \label{eq:resolv2}
\end{equation}
where $\partial_i$ is the set of vertices neighbouring $i$, and 
$\partial_i \setminus j$ denotes the set of all neighbours
of $i$ except for node $j$.  In appendix \ref{app:treeStruct} we report the
essential steps in deriving these
resolvent equations. 

The resolvent equations (\ref{eq:resolv2}) have appeared in the
literature for the first time, as far as we know,
in the study of electron localization
\cite{Abou}. Similar equations have been derived in the context of L\'evy
matrices \cite{Ciz} and in the more general context of sparse
random graphs \cite{Rogers1}.  The resolvent
equations (\ref{eq:resolv2}) form a set of $2|E|$ equations in $2|E|$-unknowns.

We interpret the set of equations for $\{  G_i^{(j)}  \}$ as a message-passing
algorithm along the
induced graph $\mathcal{G}$, similar to belief propagation which is widely used
in statistical inference and information theory \cite{Pearl}.  Here the
messages are the diagonal elements of the resolvent of the cavity graph
\cite{Rogers1}. 

Next, we extend the resolvent equations (\ref{eq:resolv2}) to Husimi
graphs with many short cycles.  
 Although the Bethe-Peierls approximation is no longer valid on these graphs,
they still contain an infinite-dimensional
character allowing for an analogous approximation.  In
statistical mechanics this forms the basis
for the Kikuchi approximation \cite{Kikuchi, Yed}. For Husimi graphs the
Kikuchi approximation gives us the resolvent equations:
\begin{eqnarray}
 G^{(i)}_{\alpha} = \left(z \bI_{\ell-1} -
\bD^{(i)}_{\alpha}- \bL_{\alpha}
-\bL^T_{\alpha}\right)^{-1} \,, \label{eq:resHus1} 
\end{eqnarray}
with $\alpha\in \partial^{(\ell-1)}_i$ and with $\partial^{(\ell-1)}_i$  the
set of all $(\ell-1)$-tuples
$(j_1, j_2, \cdots, j_{\ell-1})$ which form a cycle of length $\ell$ with the
node $i$.  The elements of the $(\ell-1)$-dimensional matrix $\bL_{\alpha}$ are
$\left[\bL_{\alpha}\right]_{nm} = \left[\bL_{(j_1, j_2, \cdots,
j_{\ell-1})}\right]_{nm} = \delta_{n(m+1)}A_{j_nj_m}$, with
$n,m=1,...,(\ell-1)$.  The quantity $\bD^{(i)}_{\alpha}$ is a diagonal matrix
with elements 
\begin{eqnarray}
 \left[\bD^{(i)}_{\alpha}\right]_{kk} = \sum_{\beta\in
\partial^{(\ell-1)}_{j_k}\setminus 
(i, j_1,\ldots, j_{k-1}, j_{k+1}, \ldots,
j_{\ell-1})
}
\bA^T_{j_k, \beta}
G^{(j_k)}_{\beta} \bA_{j_k, \beta} \,,
\nonumber 
\end{eqnarray}
with $\alpha=(j_1, j_2, \cdots, j_{\ell-1})$. 
We have also defined the $(\ell-1)$-dimensional
vector $\bA^{T}_{j_k, \beta} = \bA_{j_k, (k_1, k_2, \cdots, k_{\ell-1})} =
(A_{j_k  k_{1}} \,\,\, 0
\,\, \dots \,\, 0 \,\,\, A_{j_k  k_{\ell-1}}) $. 
The diagonal elements of the resolvent are given by
\begin{eqnarray}
 G_i = \frac{1}{z - \sum_{\alpha\in \partial^{(\ell-1)}_i} \bA^T_{i, \alpha}
G^{(i)}_{\alpha}\bA_{i, \alpha}}. \label{eq:resHus2}
\end{eqnarray}
In appendix \ref{app:cactusStruct} we elaborate on the precise derivation of
these equations.  

The resolvent equations (\ref{eq:resHus1}) determine the spectra of graphs with
many short cycles of fixed length $\ell$.  Note that they  
can be straightforwardly generalized to
graphs with variable cycle lengths and, even more general, to graphs composed
from arbitrary figures, by replacing $\bL_{\alpha}$ by the corresponding 
adjacency matrix of the figure $\alpha$ in the absence of a node
$i$, and $\bA_{j_k, \beta}$ by the submatrix of $\bA$
connecting the figure $\beta$ and the node $j_k$. In our case, the
figures of the graph are the $(\ell-1)$-tuples $\alpha$. The set of resolvent
equations (\ref{eq:resHus1})-(\ref{eq:resHus2}) can  also
be seen as a message-passing algorithm between regions of the
graph, which is similar to the generalized belief propagation
algorithm in information theory \cite{Yed}. 
Here the messages of the algorithm are the 
$(\ell-1)\times (\ell-1)$ matrices $\bG^{(i)}_{\alpha}$ sent from the
$(\ell-1)$-tuples $\alpha$ to a vertex $i$, with whom the tuple $\alpha$ forms a
cycle of length $\ell$ in the graph.  

\begin{figure}[h]
\centering
\includegraphics[scale = 0.27, angle = -90]{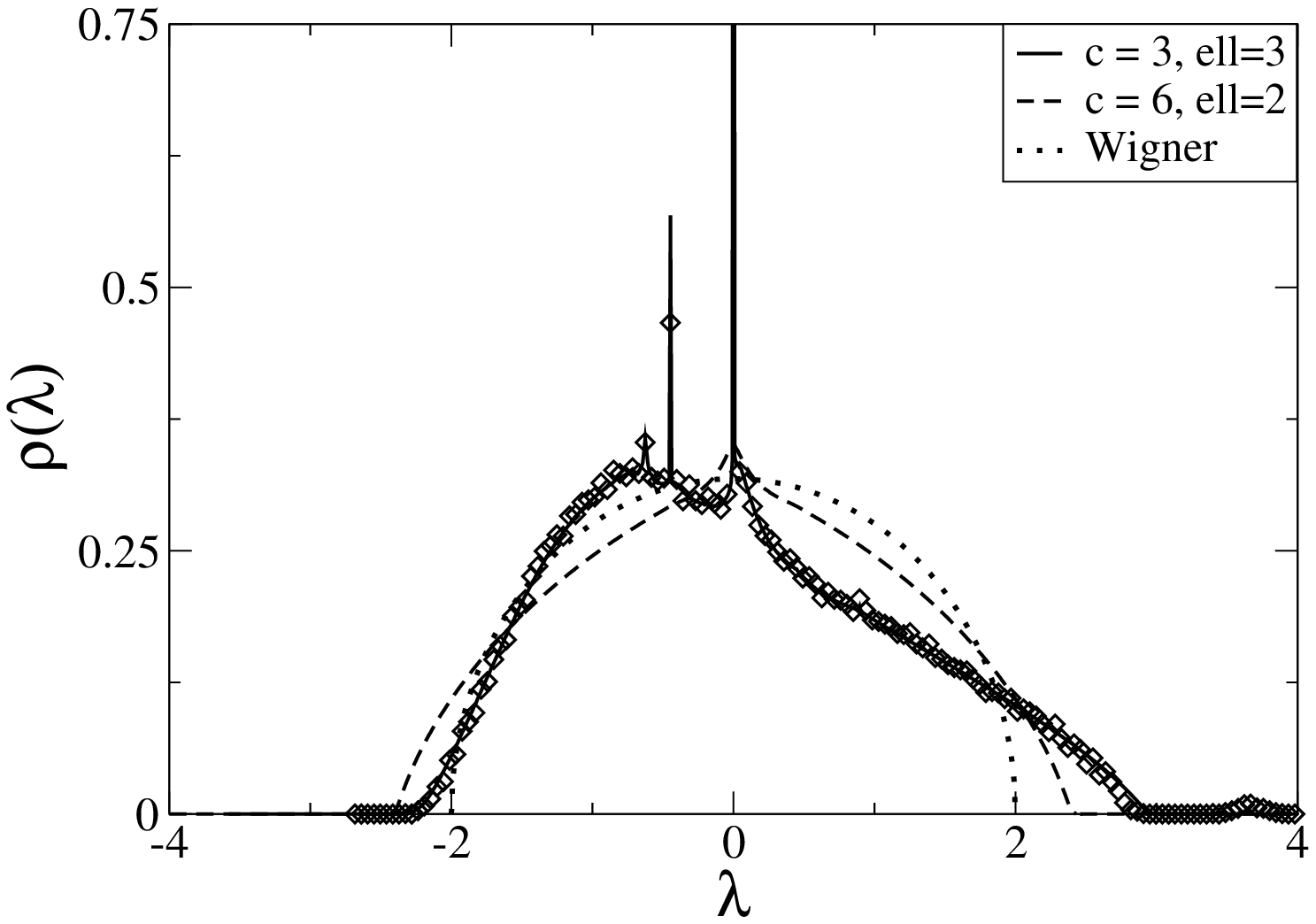}
\caption{The density of eigenvalues $\lambda$ for an
irregular triangular Husimi graph $(\ell=3)$ of mean connectivity $c=3$ with
scaling $A_{ij}\rightarrow
A_{ij}/\sqrt{2c-1}$ (solid lines) is compared with the eigenvalue
density of an irregular Erd\"os-R\'enyi graph $(\ell=2)$ of mean connectivity
$c=6$ with scaling  $A_{ij}\rightarrow
A_{ij}/\sqrt{c-1}$ (dashed lines).  Both ensembles have the same mean vertex degree.  
Results from solving the equations (\ref{eq:resHus1}-\ref{eq:resHus2}) for
$\ell=3$ with  a regulator
$\rm{Im}(z) = 10^{-4}$  are compared with direct diagonalization
results (markers).  The dotted line is the Wigner semi-circle law.} 
\label{fig:c6}
\end{figure}

We have verified the exactness of the resolvent equations
(\ref{eq:resHus1})-(\ref{eq:resHus2}) on irregular
Husimi graphs with $\ell=3$ through a comparison with direct diagonalization
results.  This is presented in figure \ref{fig:c6}, where we also compare 
the spectrum of irregular Husimi graphs with mean cycle connectivity $c=3$ 
with the spectrum of Erd\"os-R\'enyi random graphs with
the same mean vertex degree $c=6$.  From these results it appears that the
spectrum of Erd\"os-R\'enyi graphs converges faster to the Wigner semicircle for
$c\rightarrow \infty$. 

Finally, we point out that the ensemble definitions in section 2 are
global by specifying $P(\bA)$.  In the derivation of the resolvent equations
(\ref{eq:resHus1})-(\ref{eq:resHus2}) these global definitions are  not 
explicitely used. Instead, our analysis  uses the typical local  
neighbourhoods of the vertices in the graph. 
Therefore, our results are valid for all graphs which have a local neighbourhood
similar to the one given by the resolvent equations.  
The connection between the global definitions and the resolvent equations can
be made explicit through the distribution of local vertex neighbourhoods, see
for instance \cite{Bordenave}. This is how we have derived the results in figure~\ref{fig:c6}.

In the next two sections we solve the resolvent equations for the case of
undirected regular Husimi graphs for which they simplify considerably. 

\subsection{The resolvent equations for regular undirected graphs}
In this section we consider graphs with a simple topology that allows to extract exact
analytical solutions from the resolvent equations 
(\ref{eq:resHus1})-(\ref{eq:resHus2}): regular Husimi graphs with
cycles. For this ensemble of random graphs 
$\left|\partial^{(\ell-1)}_i\right|=c$ and $A_{ij} = 1$ for all connected pairs
$(i,j)$.  In figure \ref{cavities1} the square undirected Husimi graph is illustrated. 

Since there is no local disorder every node has the same local neighbourhood.  
For $N\rightarrow \infty$ the graph becomes transitive and we can set
\begin{equation}
G_{\alpha}^{(i)}(z) = \bC(z), \qquad G_{i}(z) = G(z)\,,
\end{equation}
in the resolvent equations (\ref{eq:resHus1}).  We get the following closed
equations in the $(\ell-1)$-dimensional matrices $\bC$: 
\begin{equation}
\bC(z) = \left(\begin{array}{cccc} 
z - (c-1) \mathcal{A}^{T} \bC(z) \mathcal{A}    &  -1 & 0 & \cdots \\
-1 & z - (c-1) \mathcal{A}^{T}  \bC(z) \mathcal{A}   &  -1 & \cdots \\
0 & -1  & \ddots &   \\
\vdots & \vdots &   & \ddots \end{array} \right )^{-1}, 
\label{reg}
\end{equation}
and the resolvent follows from
\begin{equation}
G(z) = \frac{1}{
z -c \:  \mathcal{A}^{T}  \bC(z) \mathcal{A} \,,
}
\end{equation}
where  $\mathcal{A}^{T} = (1 \,\,\, 0\,\, \dots \,\, 0 \,\,\, 1) $.  
It is convenient to introduce 
the scalar variable $G_s (z) = \mathcal{A}^{T}  \bC(z) \mathcal{A}$
and rewrite (\ref{reg}) as
\begin{equation}
 G_s(z) = \mathcal{A}^T \left(\begin{array}{cccc} 
z - (c-1) G_s (z)   &  -1 & 0 & \cdots \\
-1 & z - (c-1) G_s(z) &  -1 & \cdots \\
0 & -1  & \ddots &   \\
\vdots & \vdots &   & \ddots \end{array} \right )^{-1} \mathcal{A}\,.
\label{reg1}
\end{equation}
The matrix  in (\ref{reg1})
has a simple structure.   We have calculated the inverse
analytically using specific methods
for tridiagonal matrices \cite{Huang97}.
This leads to our final formula for $G_s$ when 
$\ell > 2$
\begin{equation}
G_s = \frac{2 n_{\ell-2} + 2}{n_{\ell-1}}. 
\label{Gs}
\end{equation}
The coefficients $n_{2}, \dots, n_{\ell-1}$ 
are complex numbers given by the
recurrence relation
\begin{equation}
n_i = n_1 n_{i-1} - n_{i-2}, \ \ i\geq 2
\label{alpha}
\end{equation}
with initial values $n_{0}=1$ and $n_1 = z - (c-1) G_s$.
Equations (\ref{Gs}) and (\ref{alpha}) are a set of very simple equations
that determine the resolvent elements of
undirected regular Husimi graphs. It is one of the
main results of our previous work \cite{Metz2011}. 

From the
expression of $G_s$ we   find straightforwardly
the Green function
\begin{equation}
G(z) = \frac{1}{
z -c \: G_s(z)},
\label{Greensimp}
\end{equation}
and the spectrum, according to eq. (\ref{eq:spectrumRec}) and (\ref{Greensimp}),
\begin{equation}
\rho(\lambda) = \frac{c}{\pi}  \frac{{\rm Im} G_s}
{\left( \lambda - c \: {\rm Re} G_s   \right)^2
+ c^2 \left( {\rm Im} G_s   \right)^{2}}.
\label{specGs}
\end{equation}

\subsection{Spectra of regular undirected graphs}
We solve the algebraic equations (\ref{Gs})-(\ref{alpha}) to find
the spectrum of $(\ell, c)$-regular undirected Husimi graphs. 
Due to their simple linear structure, these recurrence relations can be
solved excactly and allow to obtain the analytical form of the polynomials 
in the variable $G_s$ for any value of $\ell>2$:
\begin{itemize}
 \item $\ell=3$ corresponds with a quadratic equation: 
\begin{eqnarray}
 G_s \left[ z - (c-1) G_s \right] - G_s = 2.  \nonumber 
\end{eqnarray}
 \item $\ell=4$ gives a cubic equation: 
\begin{eqnarray}
 G_s = \frac{2 \left[ z - (c-1) G_s  \right]}{ \left[ z - (c-1) G_s  \right]^2
-2}.  \nonumber 
\end{eqnarray}
\item $\ell=5$ is determined by a quintic equation:
\begin{eqnarray}
G_s = \frac{2 \left[ z - (c-1) G_s  \right]^{3} - 4\left[ z - (c-1) G_s  \right]
+2}
{\left[ z - (c-1) G_s  \right]^{4} - 3 \left[ z - (c-1) G_s  \right]^{2} + 1}. 
\nonumber 
\end{eqnarray}
\item $\ell=6$ follows from a quartic equation:
\begin{eqnarray}
G_s = \frac{2 \left[ z - (c-1) G_s  \right]^{2} - 4}
{\left[ z - (c-1) G_s  \right]^{3} - 3\left[ z - (c-1) G_s  \right]}. 
\nonumber 
\end{eqnarray}
\item For $\ell > 6$, the polynomials have a degree larger 
than four.  
\end{itemize}
The root of these polynomial equations which gives the  expression for the
spectrum is a stable fixed point of eq. (\ref{reg1}).
In general, one  finds the roots of
polynomials in terms of radicals up to degree four. For 
larger degrees, algebraic solutions in terms of radicals 
no longer exist, apart from some particular 
situations \cite{Bruce}. The roots of general polynomials
with degree larger than four are given
in terms of elliptic functions \cite{Bruce}.  

We have solved the above equations
for $\ell=3,4,6$ and have found the analytical
expression for $\rho_{\ell}(\lambda)$, generalizing the
Kesten-McKay law \cite{Mckay, Kesten} to regular graphs with cycles.
For other values of $\ell$, eqs. (\ref{Gs})-(\ref{alpha}) are solved numerically
in a 
straightforward way, allowing to study the spectrum
as a function of $\ell$ with high accuracy.   

From the solution of the resolvent equations, we have found the
expressions for the spectra $\rho_\ell (\lambda)$ of $(\ell,c)$-regular
undirected Husimi graphs.  We point out that for $\ell=2$ we recover a
sparse regular graph without 
short cycles. In this case we have the Kesten-Mckay eigenvalue-density
distribution \cite{Mckay, Kesten}
\begin{eqnarray}
 \rho_{2}(\lambda) &=&
\frac{c}{2\pi}\frac{\sqrt{4(c-1)-\lambda^2}}{c^2-\lambda^2}, \label{eq:KM}
\end{eqnarray}
for $\lambda\in[-2\sqrt{c-1}, 2\sqrt{c-1}]$, and $\rho_2(\lambda) = 0$ otherwise. The
expression (\ref{eq:KM}) follows from solving eq. (\ref{reg1}).

For $\ell=3$ we recover the spectra $\rho_3(\lambda)$ of triangular 
Husimi graphs \cite{Eck05} 
\begin{eqnarray}
 \rho_3(\lambda) &=&
\frac{
2(c-1)}{\pi
c}\frac{\sqrt{D_3(\lambda)}}{
D_3(\lambda)+
\left[\lambda\frac{2(c-1)}{c} + (1-\lambda)\right]^2}
+\frac{1}{3}\delta_{c,2}\:\delta(\lambda; -2), \nonumber
\end{eqnarray}
for $D_3(\lambda) = 8(c-1)-(\lambda-1)^2 > 0$, and
$\rho_3(\lambda) = 0$ otherwise.
Therefore, the support of $\rho_3(\lambda)$ is  
$\lambda\in[1-2\sqrt{2(c-1)},1+2\sqrt{2(c-1)}]$.

For $\ell=4$ we find the spectrum of a square Husimi graph
\cite{Metz2011}
\begin{equation}
\rho_4(\lambda) = \frac{6\sqrt{3} \, c \, (c-1) \, q_{-}(\lambda)}
{ \pi \Big[ 2 (c-3) \lambda + c \, q_{+}(\lambda)  \Big]^{2} 
+ 3 \, \pi \, c^{2} \, q_{-}^{2}(\lambda),
} \nonumber
\end{equation}
for $D_4(\lambda)>0$ and $\rho_4(\lambda) = 0$ otherwise, where
%
\begin{equation}
q_{\pm}(\lambda) = s^{1/3}_{+} \pm s^{1/3}_{-}, 
\,\,\,\,\,
s_{\pm}(\lambda) = 9\lambda (c+1) - \lambda^{3} \pm 9
\sqrt{D_4(\lambda)},\nonumber
\end{equation}
and 
\begin{eqnarray}
 D_4(\lambda) = - \frac{2}{3}\lambda^{4} - \frac{\lambda^{2} }{3}
\left(c^2-22c+13  \right)
+ \frac{8}{3} (c-2)^{3}. \nonumber
\end{eqnarray}
The edges of $\rho_4(\lambda)$ are determined by
finding the roots of $D_4(\lambda)=0$.  For $c \rightarrow
\infty$, $\rho_4(\lambda)$ converges to the Wigner semicircle law. For $c=2$,
the 
spectrum contains a power-law
divergence $\rho_4(\lambda) \sim |\lambda|^{-1/3}$
as $|\lambda| \rightarrow 0$.   

We have also found the analytical expression for $\ell=6$:
\begin{eqnarray}
\rho_6(\lambda) = 
\frac{4 c \sqrt{R(\lambda) F(\lambda) }}
{\pi R(\lambda) \left[ \frac{(c-4)}{(c-1)} |\lambda| + 2 c R(\lambda)   
\right]^2    
+ \pi c^2 F(\lambda)
},
\nonumber 
\end{eqnarray}
for $D_6(\lambda)>0$ and $\rho_6(\lambda)=0$ otherwise, with 
\begin{eqnarray}
u_R(\lambda) &= &\left(r(\lambda) + \sqrt{D_6(\lambda)} \right)^{\frac{1}{3}} +
\left(r(\lambda) - \sqrt{D_6(\lambda)} \right)^{\frac{1}{3}}
+ \frac{\left[2c - 5 + 3 \lambda^2\right]}{3(c-1)^2}, \nonumber \\
R(\lambda) &=&  \sqrt{u_{R}(\lambda) - \frac{3 \lambda^2}{4 (c-1)^2} -
\frac{(2c-5)}{(c-1)^2} },  \nonumber \\
F(\lambda) &=& 
-\frac{|\lambda|^3}{(c-1)^3} + \frac{4(2c+1)|\lambda|}{(c-1)^3} + 4 R(\lambda)
\left[   \frac{(2c-5)}{(c-1)^2} - \frac{3 \lambda^2}{2 (c-1)^2} + u_R(\lambda)
\right],\nonumber
 \end{eqnarray}
and 
\begin{eqnarray}
D_6(\lambda) &=& q^{3}(\lambda) + r^{2}(\lambda), \nonumber \\
 q(\lambda)\:(c-1)^4 &=& \frac{-9\lambda^2 + 48(c-1) -
(2c-5)^2}{9},
\nonumber  \\
 r(\lambda)\:(c-1)^5  &=& \left[ \frac{(4c-7)^2-(4c-7)(2c-5)}{2(c-1)} +
\frac{(2c-5)^2}{3(c-1)} + \frac{(34 - 16c)}{3}\right]
\lambda^2
\nonumber \\
&&
  + \frac{16(2c-5)}{3}+ \frac{(2c-5)^3}{27(c-1)} . \nonumber
\end{eqnarray}
\begin{figure}[h]
\centering
\includegraphics[scale = 0.27, angle=-90]{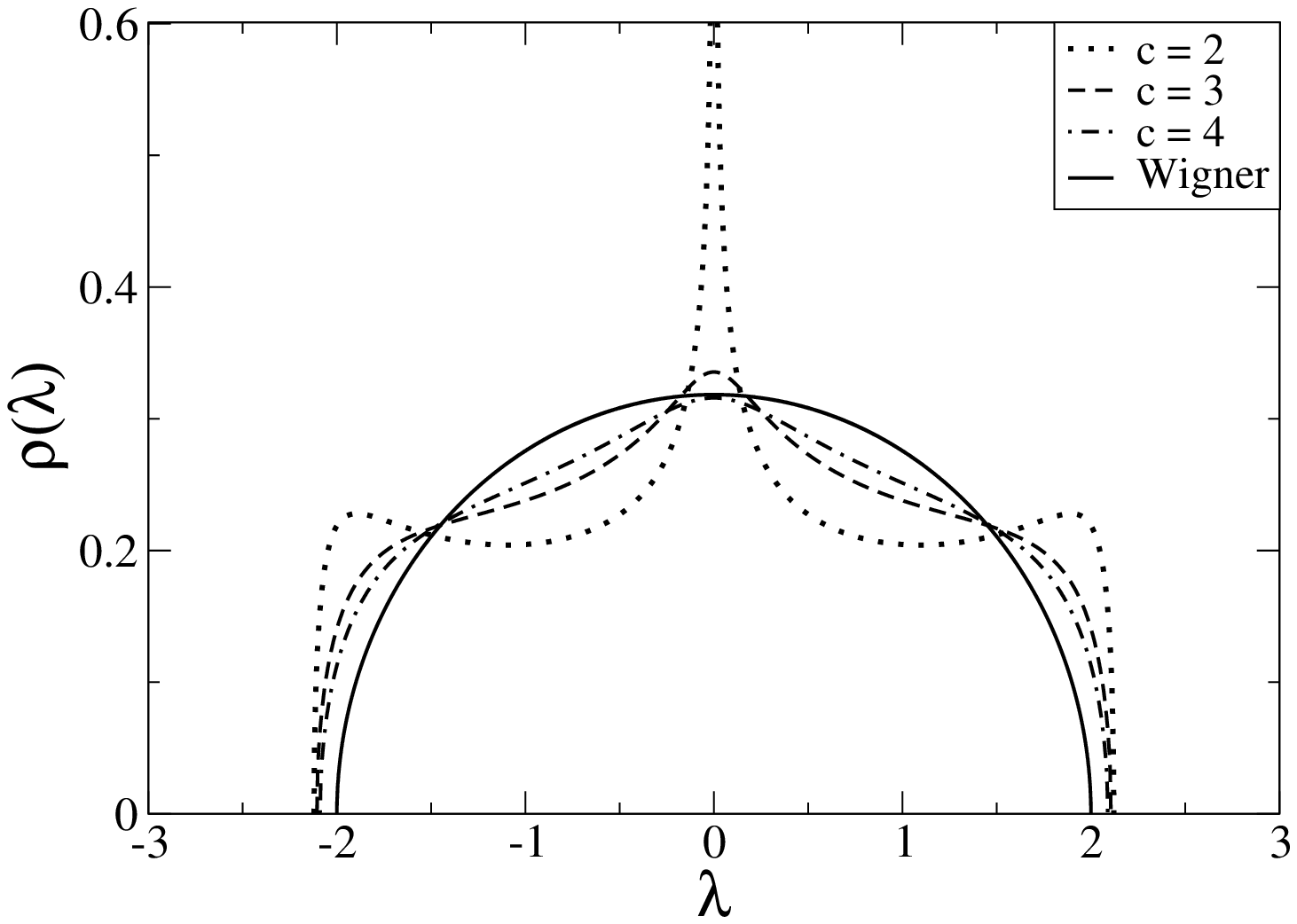}
\caption{Spectra of $(4, c)$-regular undirected Husimi graphs of 
connectivity $c=2,3,4$ and with a scaling $A_{ij} \rightarrow
A_{ij}/\sqrt{2c-1}$.
The spectrum converges to the Wigner semi-circle law 
for $c\rightarrow\infty$. 
}
\label{fig:4ell}
\end{figure}
\begin{figure}[h]
\centering
\includegraphics[scale = 0.27, angle=-90]{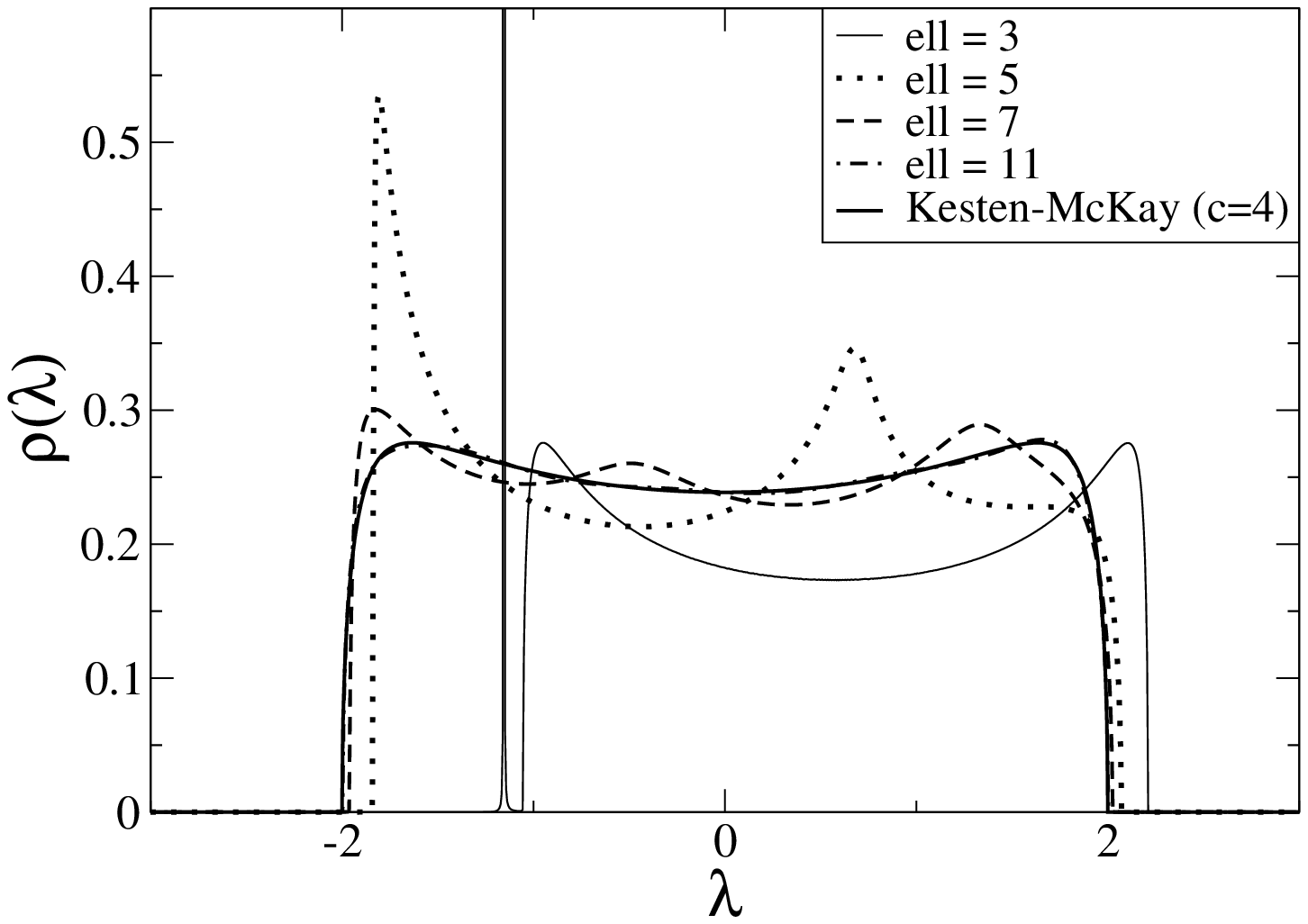}
\caption{Spectra of $(\ell, 2)$ regular 
undirected Husimi graphs with a scaling
$A_{ij} \rightarrow A_{ij}/\sqrt{2c-1}$ and with  
cycle lengths $\ell=3,5,7,11$. The spectrum converges to
the Kesten-McKay law for $\ell \rightarrow \infty$.}
\label{oddell}
\end{figure}
The function $D_6(\lambda)$ is the discriminant of the
cubic polynomial associated to the quartic polynomial for $G_s$.
The edges of $\rho_6(\lambda)$ are obtained from
the roots of $D_6(\lambda) = 0$.
The spectrum converges
to the Wigner semicircle law when $c \rightarrow \infty$.

For $\ell=5$ and $\ell\geq 7$ we obtain the spectrum
by solving numerically the set of eqs.~(\ref{Gs})-(\ref{alpha}).  
One can show that $\rho_{\ell}(\lambda)$ converges
to the Wigner law in the limit $c \rightarrow \infty$, and 
converges to the Kesten-Mckay law in the limit $\ell\rightarrow
\infty$ (this limit corresponds to a graph without short cycles).

In figure \ref{fig:4ell} we present the evolution of the spectrum as a function
of the connectivity $c$ for the case $\ell=4$.  We notice indeed the fast
convergence to
the Wigner semi-circle law for $c\rightarrow \infty$.  In figure \ref{oddell} we
present the evolution of the spectrum as a function of the cycle length $\ell$, 
for fixed $c=2$.  We see how the spectrum
converges rapidly to the Kesten-McKay law for $\ell\rightarrow \infty$.

\section{Spectral analysis of directed graphs with cycles}
In our study of  spectral properties of regular directed
graphs with cycles we follow again four
steps. First, we  write the spectrum in statistical mechanics terms \cite{Rogers2}.  This is done 
by mapping the resolvent calculation of a directed graph on a resolvent
calculation of a related undirected graph, using the Hermitization procedure \cite{FZ97}.
The resolvent of this undirected graph is then analyzed
using the methodology presented in section 3. Thereby, the infinite-dimensional
nature of the resultant graphs again allows 
to derive a closed expression in submatrices of the resolvent matrix. 
Thirdly, we show how this closed set of equations determines the spectrum of
directed regular Husimi graphs and we derive their explicit form in this case.
Finally, the resultant algebraic expressions are solved to arrive at explicit analytical
expressions in the support of the spectra.

\subsection{Statistical mechanics formulation of the spectrum}
The eigenvalues $\lambda_{1},\dots, \lambda_{N}$ of the $N \times N$ adjacency
matrix $\bA$
of directed Husimi graphs are distributed
in the complex plane, contrary to the real eigenvalues for undirected Husimi
graphs. By introducing $\partial^{*} = \frac{1}{2} \left( 
\frac{\partial}{\partial x} + i  
\frac{\partial}{\partial y} \right)$ and using the
relation $\partial^{*} (x+iy)^{-1} = \pi \delta(x) \delta(y)$, the density 
of states $\rho(\lambda)= N^{-1} \sum_{\mu=1}^{N} \delta(x - {\rm Re}
\lambda_{\mu})
\delta(y - {\rm Im} \lambda_{\mu})$ at a certain point
$\lambda=x+ i y$ can be written formally
as $\rho(\lambda) = \lim_{N\rightarrow \infty}(N \pi)^{-1} \partial^{*} {\rm Tr}
\bG(\lambda)$, where $\bG(\lambda)$ is the resolvent
$\bG(\lambda) = (\lambda - \bA)^{-1}$.
The operation $(\cdot)^{*}$ denotes complex conjugation.
The $N \times N$ resolvent $\bG(\lambda)$ is the central object
of interest and its non-analytic behavior 
at the eigenvalues of $\bA$ poses difficulties in
applying various techniques well-developed for Hermitian
matrices \cite{FZ97}. An elegant way to avoid this
problem is the Hermitization method. We define a $2N \times 2N$ 
block matrix 
\begin{equation}
\bH_{\epsilon}(\lambda) =
\left(\begin{array}{cc} 
\epsilon \bI_{N} \hfill & -i(\lambda-\bA) \\
-i(\lambda^{*}-\bA^{T}) & \epsilon \bI_{N} \hfill \end{array} \right ) \,,
\label{defH}
\end{equation}
where $\epsilon > 0$ is a regularizer
and $\bI_N$ is the $N$-dimensional identity matrix. 
The lower-left block of $\lim_{\epsilon \rightarrow
0^{+}} \bH^{-1}_{\epsilon}(\lambda)$
is precisely the matrix $\bG(\lambda)$. Thus,
the problem reduces to calculating the diagonal 
matrix elements $\mathcal{G}_{j}(\lambda, \epsilon) = 
\left[ \bH^{-1}_{\epsilon}(\lambda) \right]_{j+N,j}$ ($j=1,\dots,N$), 
from which the spectrum is determined according to 
\begin{equation}
\rho(\lambda) = - \frac{i}{N \pi} \lim_{ \epsilon \rightarrow 0^{+}, N\rightarrow \infty  }
\sum_{j=1}^{N} 
\partial^{*} \mathcal{G}_{j}(\lambda,\epsilon) \,.
\label{spectrGj}
\end{equation}

The form of the enlarged matrix $\bH$ 
depends on the problem at hand and different 
proposals have appeared in the literature
\cite{FZ97,Chalker97,Chalker98,Janik99}.
The form (\ref{defH}) is particularly convenient here, since 
its Hermitian part is a positive-definite
matrix and one can represent $\bH^{-1}$ as a 
Gaussian integral. For this purpose we
introduce a set of two-dimensional column vectors $\{ \bPhi_{i}
\}_{i=1,\dots,N}$ with complex  elements and the ``Hamiltonian'' function
\begin{equation}
\mathcal{H}(\{\bPhi_{j},\bPhi_{j}^{\dagger}\}; \lambda) =
\sum_{i=1}^{N} \bPhi_{i}^{\dagger}. \bS_{\epsilon}(\lambda)   
\bPhi_{i} + i \sum_{i,j=1}^{N} 
\bPhi_{i}^{\dagger}. \mathcal{J}_{ij} \bPhi_{j} \,,
\label{Hamilt}
\end{equation}
where
\begin{eqnarray}
\mathcal{J}_{ij} = \left(\begin{array}{cc} 0 & A_{ij} \\
A_{ji} & 0  \end{array} \right ), \,\,
\label{int2x2}
\end{eqnarray}
and $\bS_{\epsilon}(\lambda) =   \left[ \epsilon \bI_{2}  - 
i \left( x \sigma_{x} - y \sigma_{y} \right)  \right]$, with
($\sigma_{x},\sigma_{y}$)
denoting Pauli matrices.  

A graphical representation using an induced graph is again useful in these
calculations. For a non-Hermitian
matrix with real entries the induced graph is directed, contrary to an
undirected graph for real symmetric matrices. Graphically the matrix
elements $A_{ij}$ correspond then with a directed edge from node $i$ to node~$j$.

Combining the representation of $\left[ \bH^{-1}_{\epsilon}(\lambda)
\right]_{ij}$
as a Gaussian integral with eq.~(\ref{Hamilt}), $\mathcal{G}_{k}(\lambda,
\epsilon)$ reads
\begin{eqnarray}
\mathcal{G}_{k}(\lambda, \epsilon) &=& 
\frac{1}{2}\int \left[ \prod_{j=1}^{N} d \bPhi_{j} d \bPhi^{\dagger}_{j} \right]
\left[ \bPhi^{\dagger}_{k}.(\sigma_{x} + i \sigma_{y})\bPhi_{k} \right] 
\mathcal{P}  (\{ \bPhi_{j},\bPhi^{\dagger}_{j}; \lambda \}) \,, 
\label{matrH}
\end{eqnarray}
where we have introduced the complex-valued measure $\prod_{j=1}^{N} \mathcal{P}  (\{ \bPhi_{j},\bPhi^{\dagger}_{j} \}; \lambda)  d \bPhi_{j} d \bPhi^{\dagger}_{j} $ 
\begin{eqnarray}
\mathcal{P}  (\{ \bPhi_{i},\bPhi^{\dagger}_{i} \}; \lambda) = \frac{ 
\exp{\left[ -\mathcal{H}(\{ \bPhi_{i},\bPhi^{\dagger}_{i} \}; \lambda)\right]} }
{  \int \left[ \prod_{j=1}^{N} d \bPhi_{j} d \bPhi^{\dagger}_{j} \right]
\exp{\left[ -\mathcal{H}(\{ \bPhi_{i},\bPhi^{\dagger}_{i} \}; \lambda)\right]}  
} \,.
\end{eqnarray}
Equation (\ref{matrH}) shows that the local marginals
$\{ \mathcal{P}  (\bPhi_{k},\bPhi^{\dagger}_{k} ;\lambda) \}_{k=1,\dots,N}$
determine
the functions $\{ \mathcal{G}_{k} \}_{k=1,\dots,N}$, from which
the spectrum follows through eq.~(\ref{spectrGj}). 

\subsection{The resolvent equations}
We use again the infinite dimensional nature of Husimi graphs to derive an
exact equation in the resolvent elements.  
Due to the sparse structure of $\bA$, the average number of nodes
in a path connecting two randomly chosen cycles scales
as $\ln N$ (see figure \ref{fig:cavities}). This fundamental property allows 
to compute $\{ \mathcal{P}  (\bPhi_{k},\bPhi^{\dagger}_{k}) \}_{k=1,\dots,N}$
using the cavity method, as demonstrated in appendix A.2. The main difference
with the undirected case
resides in the ``state variables'' describing the nodes 
of the graph and the corresponding ``Hamiltonian''.
While in the undirected case they are scalar real variables, here
there is a two-dimensional complex-vector $\bPhi_{i}$ associated to each
vertex and they mutually interact through $\mathcal{J}_{ij}$, see
eq.~(\ref{Hamilt}).
Hence the resolvent equations involve  two-dimensional
matrices. 

Extending the cavity formulation \cite{Rogers2} to directed graphs with
cycles, see appendix A.2, we have derived the following
equation 
\begin{equation}
\mathcal{G}_{i}(\lambda,\epsilon) = \left( \bS_{\epsilon}(\lambda)     
+ \sum_{\alpha \in \partial^{(\ell-1)}_i }  \mathbb{ A}^{T}_{i \alpha} \mathcal{
D}_{\alpha}^{(i)} \mathbb{A}_{i \alpha} 
\right)^{-1}_{21}
\,,
\label{Gj}
\end{equation}
where $\partial^{(\ell-1)}_i$ is the set of all $(\ell-1)$ tuples
$(j_1, j_2, \cdots, j_{\ell-1})$ which form a cycle of length $\ell$ with 
node $i$.
The $2 \times 2(\ell-1)$ block matrix 
$\mathbb{ A}^{T}_{i \alpha} =  (\mathcal{J}_{i  j_{1}} \,\,\, \bzero
\,\, \dots \,\, \bzero \,\,\, \mathcal{J}_{i  j_{\ell-1}}) $ encodes 
the interaction between $i$ and a given
tuplet $\alpha = (j_1, j_2, \cdots, j_{\ell-1})$. 
The matrix $\bzero$ is a two-dimensional matrix filled with
zeros.
The $2(\ell-1) \times 2(\ell-1)$ block
matrices $\mathcal{ D}_{\alpha}^{(i)} =\mathcal{ D}_{(j_1, j_2, \cdots,
j_{\ell-1})}^{(i)}  $ 
fulfill the cavity equations
\begin{equation}
\mathcal{ D}_{\alpha}^{(i)} = \left( \bS_{\epsilon}(\lambda)   \otimes
\bI_{\ell-1} + \mathcal{B}_{\alpha}^{(i)} + i \mathbb{L}_{\alpha} + i
\mathbb{L}_{\alpha}^{T} \right)^{-1},
\label{cavD}
\end{equation}
where $i=1,\dots,N$ and $\alpha \in \partial^{(\ell-1)}_i$.
The matrix $\mathbb{L}_{\alpha}$ is composed of $2 \times 2$ 
block elements defined by $[\mathbb{L}_{\alpha}]_{nm} = [\mathbb{L}_{(j_1, j_2,
\cdots, j_{\ell-1})}]_{nm} 
= \delta_{n,m+1} \, \mathcal{J}_{j_{n},j_{m}}$, where $n = 2,\dots, \ell-1$.
The matrix $\mathcal{B}_{\alpha}^{(i)}$  is a diagonal
matrix formed by the following $2 \times 2$ block elements
\begin{equation}
[\mathcal{ B}_{\alpha}^{(i)} ]_{kk} = \sum_{\beta \in 
\partial^{(\ell-1)}_{j_k}\setminus 
(i, j_1,\ldots, j_{k-1}, j_{k+1}, \ldots,
j_{\ell-1})  } \mathbb{ A}^{T}_{j_{k}
\beta}  \mathcal{ D}_{\beta}^{(j_k)} \mathbb{A}_{j_k \beta},
\label{matrB}
\end{equation}
with $k=1,\dots,\ell-1$ and $\alpha = (j_1,j_2, \cdots, j_{\ell-1})$.

Once eqs.~(\ref{cavD}) have been solved, the spectrum 
follows from eqs.~(\ref{Gj}) and (\ref{spectrGj}).
The cavity equations have an interpretation 
in terms of a message-passing algorithm: the
matrix $\mathcal{ D}_{\alpha}^{(i)}$ is seen
as the message sent by the $(\ell-1)$ nodes
of cycle $\alpha$ to node $i$ of the
same cycle \cite{Yed}. This completes
the general solution of the problem. 

\subsection{The resolvent equations for regular directed graphs}
We determine now the resolvent equations for the spectrum $\rho_{\ell}$ of
infinitely large $(\ell, c)$-regular directed Husimi graphs.  These graphs have 
$|\partial^{(\ell-1)}_{i}| = c$ for $i=1,\dots,N$, i.e., each vertex is incident
to $c$ cycles of length $\ell$. We set $A_{ij}=1$ and
$A_{ji} = 0$ when there is a directed edge from node $i$
to $j$, such that the corresponding matrix $\mathcal{J}_{ij}$ 
assumes the form $\mathcal{J} = \frac{1}{2}(\sigma_{x} + 
i \sigma_{y})$. As a consequence, the matrices
$\{ \mathcal{ D}_{\beta}^{(i)}, \mathbb{A}_{i \beta}, \mathbb{L}_{\beta} \}$
become independent of the indices ($i,\beta$). It is
convenient to define the two-dimensional matrix $\bG_{A} = \mathbb{A}^{T}
\mathcal{ D} \mathbb{A}$, where $\mathbb{A}^{T} = (\mathcal{J} \,\,\, \bzero
\,\, \dots \,\, \bzero \,\,\, \mathcal{J}^{T}) $. We  write $\rho(\lambda)$ 
in terms of $\bG_{A}$ as follows
\begin{equation}
\rho_{\ell}(\lambda) = \frac{1}{i \pi} \lim_{\epsilon \rightarrow 0} 
\partial^{*} \left[ \bS_{\epsilon}(\lambda)  + c \, \bG_{A}
\right]_{21}^{-1}\,.
\label{spectrcav}
\end{equation}
From eqs.~(\ref{cavD}) and (\ref{matrB}) one
obtains that, for $\ell > 2$, the two-dimensional matrix $\bG_{A}$ solves the
equation
%
\begin{equation}
\bG_{A} = \mathbb{A}^{T} \Big [ \big( \bS_{\epsilon}(\lambda) + (c-1)
\bG_{A} \big) \otimes \bI_{l-1} 
+ i \mathcal{J} \otimes \bL_{\ell-1} + i \mathcal{J}^{T} \otimes
\bL_{\ell-1}^{T}  \Big]^{-1} \mathbb{A}  \,,
\label{eqD}
\end{equation}
where $\bL_{\ell-1}$ is a $(\ell-1)$-dimensional matrix
with elements $[\bL_{\ell-1}]_{i j} = \delta_{i,j+1}$.  
The derivative of eq.~(\ref{eqD})
yields an equation in $\partial^{*}\bG_{A}$, which has to be 
solved together with (\ref{eqD}) to find $\rho_{\ell}(\lambda)$ through 
eq.~(\ref{spectrcav}). 

\begin{figure}[h]
\centering
\subfigure[ $\ell = 3$]{
\includegraphics[scale=0.48]{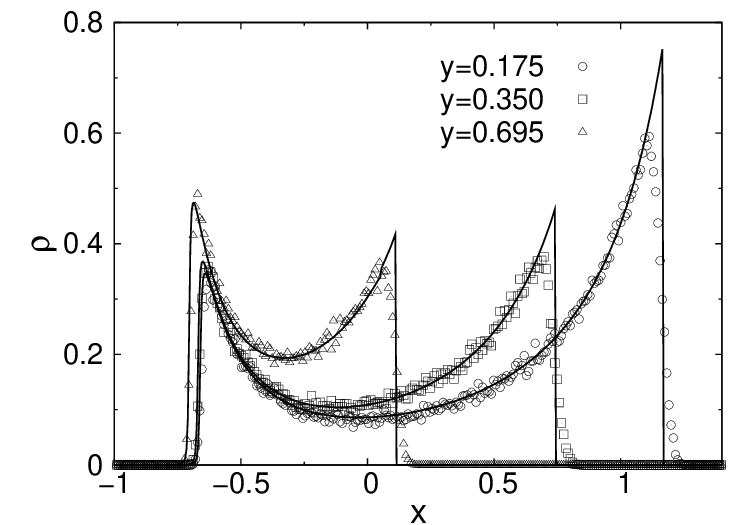}}
\subfigure[ $\ell = 4$]{
\includegraphics[scale=0.48]{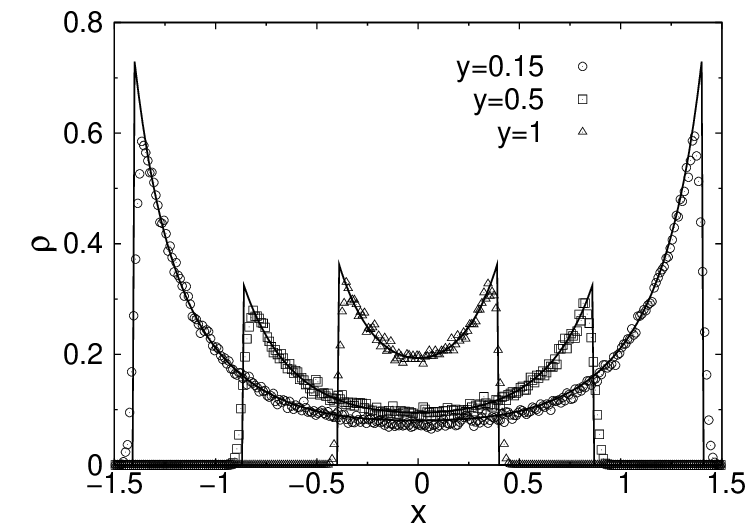}}
\caption{Three cuts of the spectrum  $\rho_\ell(\lambda)$ of $(\ell, c)$-regular
directed Husimi graphs along
the real direction for $c=2$, $\ell=3$ \cite{Metz2011} and $c=2$, $\ell=4$.
These results (solid lines) are obtained from
the numerical solution of eq.~(\ref{eqD}) and they are compared with 
direct diagonalization results (markers) for an
ensemble of $3 \times 10^{4}$ matrices of dimension
$N = 10^{3}$.
}
\label{cuts}
\end{figure}

Equation (\ref{eqD}) 
allows to derive accurate numerical results
for the spectrum of directed Husimi graphs 
as a function of $\ell$. As an illustration, we present in figure \ref{cuts}
some cuts of $\rho_3(\lambda)$ and $\rho_4(\lambda)$ along
the real direction for fixed values of $y$.
These results correspond very well with direct 
diagonalization, confirming the exactness of (\ref{eqD}).
For a three-dimensional graph of $\rho_{3}(\lambda)$
we refer the reader to \cite{Metz2011}. 
Analogously to undirected
Husimi graphs, $\rho_{\ell}(\lambda)$ converges
to the spectrum of a directed regular graph without short cycles for $\ell
\rightarrow \infty$ \cite{BordChaf, Neri2012, RogerThesis}:
\begin{equation}
\rho_{\infty}(\lambda)= \frac{c-1}{\pi}  \left( \frac{c }{c^2 -|\lambda|^{2}}
\right)^{2}\,,
\label{spectruml2}
\end{equation}
for $|\lambda|^{2} < c $, and $\rho_{\infty}(\lambda) = 0$ otherwise.  This convergence is shown 
numerically in \cite{Metz2011}.
By rescaling $A_{ij} \rightarrow A_{ij}/\sqrt{c-1}$, the 
solution of (\ref{eqD})  leads
to Girko's law in the limit $c \rightarrow \infty$.  

\subsection{The spectral boundaries for regular directed graphs}

In order to derive analytical equations for the  support of
$\rho_{\ell}(\lambda)$,
we determine the  inverse of the $2(\ell-1) \times 2(\ell-1)$ matrix 
present in eq.~(\ref{eqD}). Since this matrix has a tridiagonal block structure
its inverse can be  computed analytically using the method  in
\cite{Huang97}. Applying this scheme to the matrix in eq.~(\ref{eqD}), we have
simplified (\ref{eqD}) into an equation involving
sums and products of only
two-dimensional matrices.  This equation forms the equivalent for directed
regular graphs of the equation (\ref{Gs}) for undirected regular graphs.   
The resultant equations in $\bG_A$ can be solved
using the following {\it ansatz} \cite{Rogers2} 
\begin{equation}
\bG_{A} = \left(\begin{array}{cc} 
a   &  i b  \\
i b^{*}  & d \end{array} \right ) \,,
\label{ansatz}
\end{equation}
where $b$ is complex and  $a$ and
$d$ are both real variables. The Hermitian
part of this matrix is positive-definite
provided that $a$ and $d$ are positive.  This
condition ensures that the Gaussian integrals
in the cavity method are convergent \cite{Rogers2}.
Solving numerically eq.~(\ref{eqD})
for a finite regularizer $\epsilon > 0$, we find that $a >0$ and
$d > 0$. In the limit $\epsilon \rightarrow 0^{+}$, $a$ and $d$ vanish
at the boundary of $\rho(\lambda)$. Therefore, setting $\epsilon \rightarrow
0^{+}$ and $a=d=0$ in eq.~(\ref{eqD}),
and solving the resulting equations for $b$ leads
to an analytical expression for the support of $\rho_\ell(\lambda)$.
In this way we obtain the following equations for $b$ and $\lambda$ at the
boundaries of the support:
\begin{itemize}
\item{$\ell = 3$: 
\begin{eqnarray}
&b& = \frac{\left[ \lambda^{*} - (c-1) b^{*}  \right]^{2}}{|\lambda - (c-1) b
|^{4}}, \label{ell3b}   \\
&|\lambda& - (c-1) b |^{4} = (c-1) \left[ 1 + |\lambda - (c-1) b |^{2}  \right].
\label{ell3polyn}
\end{eqnarray}
}
\item{$\ell = 4$: 
\begin{eqnarray}
&b& = \frac{\left[ \lambda^{*} - (c-1) b^{*}  \right]^{2}}{|\lambda - (c-1) b
|^{6}}, \label{ell4b}  \\
&|\lambda& - (c-1) b |^{6} = (c-1) \left[ 1 + |\lambda - (c-1) b |^{2} +
|\lambda - (c-1) b |^{4}  \right].
\label{ell4polyn}
\end{eqnarray}
}
\end{itemize}
The solutions of the polynomials (\ref{ell3polyn}) and (\ref{ell4polyn}) 
in the variable $s \equiv |\lambda - (c-1) b |$ are given by
\begin{itemize}
\item{$\ell = 3$: 
\begin{equation}
s = \left[ \frac{(c-1)}{2} \left( 1 + \sqrt{1 + \frac{4}{(c-1)}}    \right)
\right]^{\frac{1}{2}} \,,
\label{rell3}
\end{equation}
}
\item{$\ell = 4$: 
\begin{eqnarray}
s = \left[ [R_{+}(c)]^{\frac{1}{3}} + [R_{-}(c)]^{\frac{1}{3}} + \frac{1}{3}
(c-1) \right]^{\frac{1}{2}} \,,
\label{rell4}
\end{eqnarray}
}
\end{itemize}
where we have defined
\begin{eqnarray}
R_{\pm}(c) &=& \frac{1}{27} (c-1)^3 + \frac{1}{3} (c-1)^2 \left(\frac{1}{2} \pm
F(c)   \right)
+ \frac{1}{2} (c-1) \,, \nonumber \\
F(c) &=& \left[ \frac{1}{81} (c-1)^2 + \frac{1}{9} (c-1) \kappa
+ \frac{1}{4} \kappa^{2}
- \frac{1}{81} (c-1)^2 \kappa^{3}
\right]^{\frac{1}{2}} \,, \nonumber
\end{eqnarray}
with $\kappa =  1 + \frac{3}{(c-1)}$.
By parametrizing $\lambda - (c-1) b = s \exp{(i t)}$, with 
$t \in [0,2 \pi]$, and substituting this form in
eqs.~(\ref{ell3b}) and (\ref{ell4b}), we find the following expressions for the
boundary of the support $\lambda_\ell(t)$
of triangular $(\ell=3)$ and square $(\ell=4)$ regular directed Husimi graphs:
\begin{itemize}
\item{$\ell = 3$: 
\begin{equation}
\lambda_3(t) = s \exp{(it)} + \frac{(c-1)}{s^{2}}  \exp{(-2it)} \,,   
\label{ell3ba}  
\end{equation}
}
\item{$\ell = 4$: 
\begin{equation}
\lambda_4(t) = s \exp{(it)} + \frac{(c-1)}{s^{3}}  \exp{(-3it)} \,.   
\label{ell4ba}  
\end{equation}
}
\end{itemize}
These are the parametric equations which describe, for each
corresponding $\ell$, an hypotrochoid in the
complex plane.  The parameter $s$ as a function of
the cycle degree $c$ for $\ell=3$ and $\ell=4$
is given by, respectively, eqs.~(\ref{rell3})
and (\ref{rell4}).

\begin{figure}[h!]
\center
\includegraphics[scale=1.1]{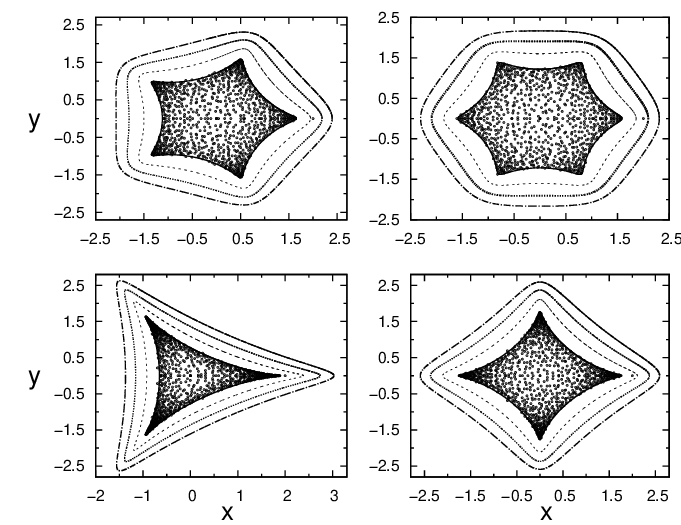}
\caption{Analytical solutions for
the boundary of the support of directed $(\ell, c)$-regular Husimi graphs for
several values of the cycle length $\ell$ and the following values $c$ of the number of
cycles incident to each vertex: $c=2$ (solid line), $c=3$ (dashed line),
$c=4$ (dotted line) and
$c=5$ (dot-dashed line). The hypotrochoids have a rotational symmetry by the angle
$2\pi/\ell$, from which one obtains the value of the cycle length $\ell$. 
Direct diagonalization
results (dots) for $N \times N$ matrices with $N=1000$ and $c=2$ are shown.}
\label{mozaic}
\end{figure}

A hypotrochoid is a cyclic function in the complex plane
which is drawn by rotating a small circle of radius $r$ in a larger circle of
radius $R$ \cite{Lawrence}.  The support of triangular and square Husimi graphs
is therefore given by
hypotrochoids with, respectively, $R/r =3$ and $R/r = 4$.  
These analytical results for the support of the spectra of directed Husimi
graphs for $\ell=3$ and $\ell=4$ are shown in the lower graphs of
figure \ref{mozaic}. The agreement
with direct diagonalization results for $c=2$
is excellent, confirming the exactness of our analytical results.
Based on the form of eqs.~(\ref{ell3b}-\ref{ell4polyn}), we
conjecture that, for a given $\ell$ and $c$, the following
equations are fulfilled at the boundary of the support of
$\rho_\ell(\lambda)$
\begin{eqnarray}
b &=& \frac{\left[ \lambda^{*} - (c-1) b^{*}  \right]^{\ell-1}}{s^{2(\ell-1)}}
\,, \label{eqbanyell}   \\
s^{2(\ell-1)} &=& (c-1) \sum_{n=0}^{\ell-2} s^{2n} \,.
\label{eqranyell}
\end{eqnarray}
Substituting $\lambda - (c-1) b = s \exp{(it)}$ 
($t \in [0,2 \pi]$) in eq.~(\ref{eqbanyell}) reads
\begin{equation}
\lambda_\ell(t) = s \exp{(it)} + \frac{(c-1)}{s^{\ell-1}} 
\exp{\left[-i(\ell-1)t\right]} \,
\label{hypogen}
\end{equation}
for the boundary of the support of a $(\ell,c)$ directed regular Husimi graph.
Remarkably, eq. (\ref{hypogen}) is a 
hypotrochoid with a fraction $R/r = \ell$.
The parameter $s$ is determined from the roots of a polynomial
of degree $\ell$ in the variable $s^{2}$, see eq.~(\ref{eqranyell}). 
Equation (\ref{eqranyell}) can
also be written as
\begin{equation}
s^2 = c - \frac{(c-1)}{s^{2(\ell-1)}} \,.
\label{simple}
\end{equation}
We have found an analytical expression for the roots
for $\ell=3$ and $\ell=4$.   
For larger
values of $\ell$, we have solved eq.~(\ref{simple}) 
numerically and, by choosing the stable solution, we have derived accurate
values for the parameters of
the hypotrochoids. We present explicit results for $\ell=5$ and $\ell=6$
in the upper graphs of figure \ref{mozaic}. Direct
diagonalization results exhibit once more
an excellent agreement with the theoretical
results, strongly supporting our conjecture that the support of the spectrum of
directed regular Husimi graphs for general $\ell$ is given
by eqs. (\ref{hypogen}-\ref{simple}). The support of regular Husimi graphs
converges to the circle $|\lambda|^{2} = c$ in the limit $l \rightarrow \infty$,
corresponding with the expression 
(\ref{spectruml2}) for a graph without short cycles.

\section{Conclusion}
In this work we have obtained the spectra of (un)directed  Husimi
graphs.  The main result is a set of exact equations which determines a
belief-propagation like algorithm in the resolvent elements of the matrix.  For
irregular graphs we have shown a very good correspondence between
direct diagonalization results and our approach.  For regular graphs we have
derived several novel analytical expressions for the spectrum of undirected
Husimi
graphs and the boundaries of the spectrum of directed Husimi graphs. 
Remarkably, the boundaries of directed regular Husimi graphs consist
of hypotrochoid functions in the complex plane.

Our results indicate that, at high connectivities, the spectrum of undirected
random
graphs converges to the Wigner semicircle law, while the spectrum of directed
random graphs converges to Girko's circular law.  This convergence seems to be
rather universal and independent of the specific graph topology. 
It would be interesting to
better understand the conditions under which finitely connected graphs
converge to these limiting laws \cite{Tao1, Wood, Dumitriu, Tran}. 

Finally, we point out that the
eigenvalues of random unistochastic
matrices are distributed over hypocycloids in the complex
plane \cite{Karol03}. This close similarity with 
our results suggests an interesting connection between 
the spectra of $\ell \times \ell$
unistochastic matrices and regular directed Husimi 
graphs with cycles of length $\ell$.
\section*{Acknowledgments}
This paper is dedicated to Fritz Gesztesy, on the occasion of his 60th
birthday.  DB wants to thank Fritz not only for many years of
stimulating and fruitful collaborations but especially for a lifetime
friendship!
FLM is indebted to Karol \.{Z}yczkowski for illuminating discussions.

\appendix
\section{Cavity method applied to random matrices} \label{appA}
We present the essential steps to determine the resolvent equations using the
cavity method \cite{Mezard, Parisi}.  This
method is based on the introduction of cavities in a graph $\mathcal{G}$ forming
subgraphs
$\mathcal{G}^{(i)}$, where the node $i$ and all of its incident edges have been
removed, see figure \ref{fig:cavities}. In
analogy one can also remove the i-th column and the i-th row from a 
matrix $\bA$ to obtain the submatrix $\bA^{(i)}$.

\subsection{Resolvent equations for locally tree-like graphs}\label{app:treeStruct}
\begin{figure}
\hfill
\begin{minipage}{.48\textwidth}
\begin{center}
\includegraphics[ width =0.8 \textwidth]{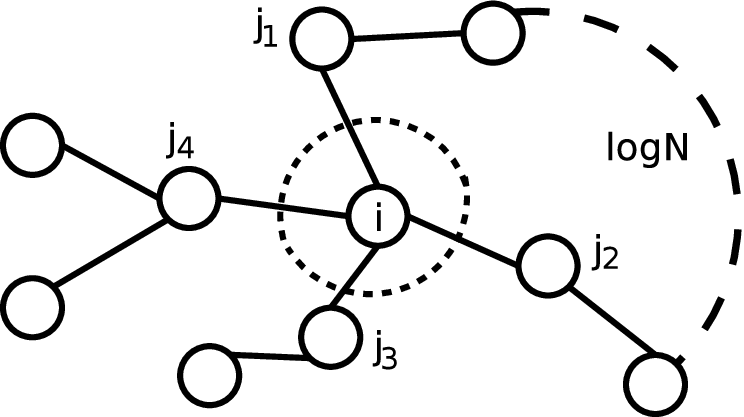}
\end{center}
\end{minipage}
\hfill
\begin{minipage}{.48\textwidth}
\begin{center}
\includegraphics[width=0.8 \textwidth]{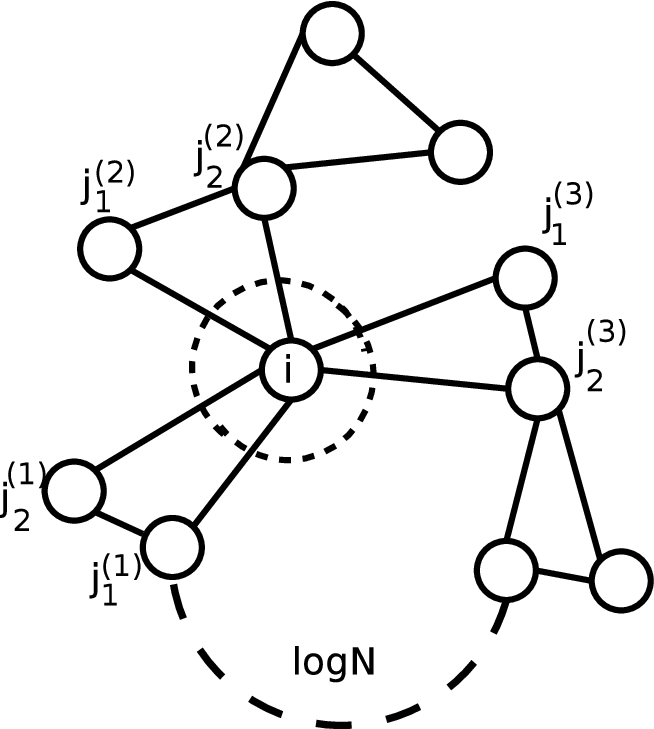}
\end{center}
\end{minipage}
\caption{Left: a graph with a locally tree-like structure.  Introducing
a cavity at the vertex $i$ decouples neighbouring vertices (in the sense that
their mutual distance scales with ${\rm log}(N)$). Right: a Husimi graph. After
introducing a cavity at the vertex $i$ neighbouring pair of nodes become
decouple.}
\label{fig:cavities}
\end{figure}

The cavity method is based on the consideration that  a
probability distribution $P(\bx;z)$
defined on a locally tree-like graph  has the factorization property: 
\begin{eqnarray}
 P^{(j)}_{\partial_j}(x_{\partial_j};z) = \prod_{k\in
\partial_j}P^{(j)}_{k}(x_k;z).
\end{eqnarray}
The quantity  $P^{(j)}_i$ is the $i$-th marginal of $P^{(j)}(\bx; z)$ on the
cavity subgraph
$\mathcal{G}^{(j)}$ of $\mathcal{G}$, and
$P^{(j)}_{\partial_j}$  is the marginal of $P^{(j)}(\bx;z)$ with respect to the
set of variables $\partial_j$.  In the language of spin models, the factorization
property follows from the locally tree-like structure of a typical neighbourhood
in the graph, see figure \ref{fig:cavities}. 

Following the derivation as presented in \cite{Rogers1}, we find a set of closed
equations in the marginals $P^{(j)}_i$  
\begin{eqnarray}
\lefteqn{ P^{(j)}_i(x_i;z) \sim \exp\left(-i\frac{zx_{i}^2}{2}\right)}&&
\label{eq:marg}\\  
&&\times \int \left(\prod_{k\in
\partial i\setminus j}dx_k P^{(i)}_k\left(x_k;z\right)\right) \exp\left(i
x_i \sum_{k\in \partial_i\setminus j }A_{ik}x_k\right), \nonumber 
\end{eqnarray}
from which the marginals $P_i$ follow: 
\begin{eqnarray}
  \lefteqn{ P_i(x_i;z) \sim
\exp\left(-i\frac{zx_{i}^2}{2}\right)}&&\label{eq:marg2}\\ 
&&\times \int \left(\prod_{k\in
\partial i}dx_k P^{(i)}_k\left(x_k;z\right)\right) \exp\left(i
x_{i} \sum_{k\in \partial_i }A_{ik}x_k\right).\nonumber
\end{eqnarray}
Finally, we use the fact that the $P^{(j)}_i$ are Gaussian functions
\begin{eqnarray}
 P^{(j)}_i(x_{i};z) = \sqrt{\frac{i}{2\pi
G^{(j)}_i(z)}}\exp\left(-i\frac{x_{i}^2}{2G^{(j)}_i(z)}\right),
\label{eq:ansatz}
\end{eqnarray}
to recover the resolvent equations (\ref{eq:resolv2}), after
substitution of (\ref{eq:ansatz}) in (\ref{eq:marg}) and (\ref{eq:marg2}).  
\subsection{Resolvent equations for Husimi graphs}\label{app:cactusStruct}
We apply now a similar  logic to graphs with many cycles which have an
infinite-dimensional
structure. In this case we  use the factorization property
\begin{eqnarray}
 P^{(i)}_{\partial^{(\ell-1)}_i}\left(x_{\partial^{(\ell-1)}_i};z\right) =
\prod_{\alpha \in
\partial^{(\ell-1)}_i}P^{(i)}_{\alpha}\left(x_{\alpha};z\right).
\end{eqnarray}
on the marginals of the distribution $P(\bx;z)$.  The factorization property
follows from the fact that the average distance between  different branches
connected to $i$ is of the order $\ln N$ after removal of $i$, see figure
\ref{fig:cavities}.
Note that 
$P^{(i)}_{\alpha}$ is the marginal of $P(\bx;z)$ for a $(\ell-1)$ tuple 
$\alpha = (j_{1},\dots,j_{\ell-1})$ forming a
cycle with the $i$-th vertex. 

We have generalized the derivation for regular
Husimi graphs presented in \cite{Metz2011}  to the
case of arbitrary Husimi graphs.   We find a set of closed equations in the
marginals
$P^{(i)}_{\alpha}$: 
\begin{eqnarray}
&& \label{eq:marginalst}\\
 \lefteqn{P^{(i)}_{\alpha}\left(x_{\alpha};z\right) \sim 
 \exp\left(-\frac{iz}{2}\sum^{\ell-1}_{k=1}x^2_{j^{(\alpha)}_k} + i
\sum^{\ell-2}_{k=1}A_{j^{(\alpha)}_k \: j^{(\alpha)}_{k+1}} \: \: 
x_{j^{(\alpha)}_k}x_{j^{(\alpha)}_{k+1}}\right) } &&\nonumber \\ 
&&\times \int \prod_{j\in \alpha} \left[\prod_{\beta \in
\partial^{(\ell-1)}_j\setminus \alpha}dx_{\beta}
P^{(j)}_{\beta}\left(x_{\beta};z\right)\right]
\nonumber \\
&& \times \prod_{j\in \alpha} \left[\prod_{\beta \in
\partial^{(\ell-1)}_j\setminus \alpha} \exp\left(i\sum_{j\in
\alpha}\sum_{\beta\in \partial^{(\ell)}_j\setminus
\alpha}\left( A_{j j^{(\beta)}_1}\:\:  x_{j^{(\beta)}_1}x_{j} +
A_{j^{(\beta)}_{\ell-1}j}\:\:  x_{j^{(\beta)}_{\ell-1}}x_{j}
\right)\right)\right],
\nonumber 
\end{eqnarray}
with the $(\ell-1)$-tuple $\alpha = \left(j^{(\alpha)}_1, j^{(\alpha)}_2,
\cdots,
j^{(\alpha)}_{\ell-1}\right)$.    The marginals $P_i$ are given as a function of
the marginals $P^{(i)}_{\alpha}$:
\begin{eqnarray}
 &&\label{eq:marginalsTT}\\
 \lefteqn{P_{i}\left(x_i;z\right) \sim 
 \exp\left(-i\frac{z x_{i}^2}{2}\right)}
&&\nonumber \\ 
&&\times \int  \prod_{\alpha \in
\partial^{(\ell-1)}_i}dx_{\alpha}
P^{(i)}_{\alpha}\left(x_{\alpha};z\right)
\prod_{\alpha}\exp\left(i\: A_{i j^{(\alpha)}_1}\: \: x_{j^{(\alpha)}_1} \,
x_{i} + 
i\:A_{j^{(\alpha)}_{\ell-1} i}\:\: x_{j^{(\alpha)}_{\ell-1}} \, x_{i} \right).
\nonumber 
\end{eqnarray}
We now use the Gaussian ansatz for the marginals
$P^{(i)}_{\alpha}\left(x_{\alpha};z\right)$: 
\begin{eqnarray}
 P^{(i)}_{\alpha}\left(x_{\alpha};z\right) \sim \exp\left(-\frac{i}{2}
\bx^T \left(\bG^{(i)}_\alpha\right)^{-1} \bx \right), 
\end{eqnarray}
with $\bG^{(i)}_\alpha$ the $(\ell-1)\times (\ell-1)$ submatrix of the
resolvent $\bG^{(i)}$ of the cavity matrix $\bA^{(i)}$: 
\begin{eqnarray}
 \bG^{(i)}_{\alpha} = \left(\begin{array}{cccc}
G^{(i)}_{j^{(\alpha)}_1j^{(\alpha)}_1} & G^{(i)}_{j^{(\alpha)}_1j^{(\alpha)}_2}
& \cdots & G^{(i)}_{j^{(\alpha)}_1j^{(\alpha)}_{\ell-1}} \\ 
G^{(i)}_{j^{(\alpha)}_2j^{(\alpha)}_1} & G^{(i)}_{j^{(\alpha)}_2j^{(\alpha)}_2}
& \cdots & G^{(i)}_{j^{(\alpha)}_2j^{(\alpha)}_{\ell-1}} \\  \vdots & \vdots &
\cdots & \vdots \\G^{(i)}_{j^{(\alpha)}_{\ell-1}j^{(\alpha)}_1}  & 
G^{(i)}_{j^{(\alpha)}_{\ell-1}j^{(\alpha)}_2} 
& \cdots & G^{(i)}_{j^{(\alpha)}_{\ell-1}j^{(\alpha)}_{\ell-1}} 
\end{array}\right).
\end{eqnarray}
We also use a Gaussian ansatz of the type (\ref{eq:ansatz}) for the marginal
$P_i$.  Substitution of these ans\"atze in  (\ref{eq:marginalst}) and
(\ref{eq:marginalsTT}) gives the resolvent equations (\ref{eq:resHus1}) and
(\ref{eq:resHus2}).

We remark that the resolvent equations
can also be derived using methods from random matrix theory ,
i.e.~by recursively
applying the Schur-complement formula \cite{Bordenave, Neri2012}. 

\bibliographystyle{amsalpha}

\end{document}